\documentclass[12pt,preprint]{aastex}

\begin{document}

\title{[\ion{Fe}{4}] Emission in Ionized Nebulae}

\author{M\'onica Rodr\'\i guez}
\affil{Instituto Nacional de Astrof\'\i sica, \'Optica y Electr\'onica
	INAOE, Apdo Postal 51 y 216, 72000 Puebla, Pue., Mexico}
\email{mrodri@inaoep.mx}

\begin{abstract}
This paper presents an analysis of [\ion{Fe}{4}] emission based on new
identifications and previous measurements of [\ion{Fe}{4}] lines in
\objectname{30~Doradus}, \objectname{IC~4846}, \objectname{M42},
\objectname{SMC~N88A}, and \objectname{SBS~0335$-$052}.
The Fe abundances obtained by adding the abundances of the relevant Fe
ions (mainly Fe$^{++}$ and Fe$^{3+}$) are found to be lower, by factors in the
range 2.6--5.9, than the Fe abundances implied by [\ion{Fe}{3}] emission and an
ionization correction factor derived from ionization models.
The most likely explanation of this discrepancy is that either the collision
strengths for [\ion{Fe}{4}] or the Fe ionization fractions predicted by models
are unreliable.
The available data neither allow one to distinguish between these two
possibilities nor to exclude another possible explanation: that the discrepancy
implies the presence of a gradient in the Fe abundance within the ionized gas.
Further measurements of [\ion{Fe}{4}] lines and checks on the Fe$^{3+}$
atomic data and ionization models are needed to reach a definitive conclusion.
The discrepancy introduces an uncertainty in the determination of Fe abundances
in ionized nebulae. This uncertainty has implications for our understanding of
both the evolution of dust in ionized nebulae and the chemical history of low
metallicity galaxies.
\end{abstract}


\section{INTRODUCTION}

The first detection of an [\ion{Fe}{4}] line in an \ion{H}{2} region is due
to \citet{rub97}, who measure [\ion{Fe}{4}]~$\lambda$2836.56
in the UV spectrum of the Orion nebula (\objectname{M42}).
From this line and two previous ionization models for Orion, \citeauthor{rub97}\
find $\mbox{Fe}/\mbox{H}$ lower, by factors of 6.5 and 19, than the value the
models need to reproduce [\ion{Fe}{2}] and [\ion{Fe}{3}] emission in
\objectname{M42}, $\mbox{Fe}/\mbox{H}=3\times10^{-6}$.
According to \citeauthor{rub97}, the difference between the results obtained
from the two models is due mostly to the different average electron temperatures
($T_{\rm e}$) predicted by each model.
Since the two Orion models and the measurement of the [\ion{Fe}{4}] line
correspond to different regions in the nebula, an underlying assumption in the
above comparison is that the gaseous Fe abundance remains roughly constant
within the ionized gas.

On the other hand, \citet{rod02} calculates the Fe abundances for 12 regions
in \objectname{M42} from the Fe$^+$ and Fe$^{++}$ abundances and ionization
correction factors (ICFs) for the contribution of Fe$^{3+}$, obtained from
grids of photoionization models.
Since the Fe$^+$ abundance is very low for all positions
($\mbox{Fe}^+/\mbox{Fe}^{++}<0.3$), the results depend
mostly on the derived Fe$^{++}$ abundances, and these are based on the atomic
data leading to the best fit between the observed and predicted relative
intensities of the [\ion{Fe}{3}] lines.
The final Fe abundances are lower than those derived in previous studies based
on older atomic data, and show variations of more than a factor of 2:
$\mbox{Fe}/\mbox{H}=0.8\mbox{--}1.8\times10^{-6}$.
Taking this into account, the discrepancy found by \citet{rub97} could be
reduced and now gets as close
as a factor of $\sim2$, and \citet{rod02} argues that given all the
uncertainties involved in the calculations, this discrepancy might not be
significant.
The main uncertainties arise from the calculation of the Fe$^{3+}$ abundance
from one weak UV line, which is very sensitive to $T_{\rm e}$ and the
extinction correction, and from the lack of measurements of [\ion{Fe}{2}],
[\ion{Fe}{3}], and diagnostic lines at exactly the same position in the
nebula.
The measurement of optical [\ion{Fe}{4}] lines would solve this difficulty,
but these lines are weak and difficult to observe.
Since Fe$^{3+}$ is an important or dominant ionization state in most \ion{H}{2}
regions, the reality of this underabundance implied by [\ion{Fe}{4}] emission,
and the reasons behind it, are critical issues in our understanding of both the
evolution of dust in \ion{H}{2} regions \citep[see][]{rod02} and the chemical
evolution at low metallicities \citep[see][]{izo99}.

This paper presents the identification and analysis of one [\ion{Fe}{4}] line
at $\lambda6739.8$ in the optical spectra of \objectname{M42} observed by
\citet{bal00}.
The same line is identified in the spectra of the planetary nebula 
\objectname{IC~4846} \citep{hyu01}. 
Upper limits to the intensities of [\ion{Fe}{4}]~$\lambda\lambda6734.4$, 6739.8
are obtained from the spectra of \objectname{30~Doradus} in the LMC observed by
\citet{ape03}.
Another optical [\ion{Fe}{4}] line at $\lambda\sim4904$, tentatively identified
by \citet*{izo01} in the spectra of the blue compact dwarf galaxy
\objectname{SBS~0335$-$052}, and the measurement of
[\ion{Fe}{4}]~$\lambda$2836.56 in \objectname{SMC~N88A} by \citet{kurt99} are
used to complete a set of data in which to perform an analysis of [\ion{Fe}{4}]
emission and the Fe abundance in different ionized nebulae.

\section{[\ion{Fe}{4}] LINES IN SEVERAL OBJECTS}

\begin{figure}
\plotone{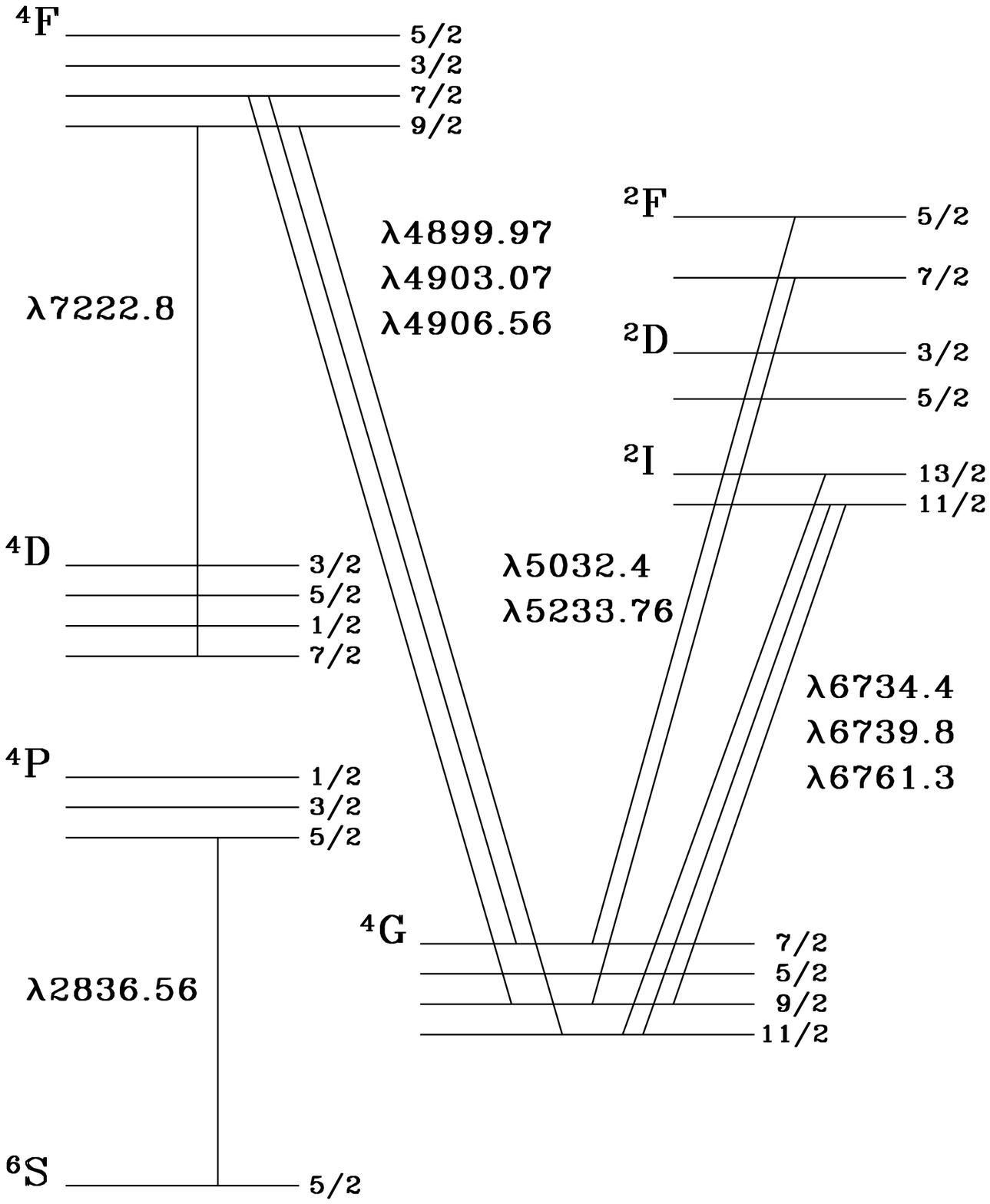}
\caption{Schematic level diagram of Fe$^{3+}$ up to 6.6 eV, with the
transitions giving rise to the lines discussed in the text. Lines are labeled
with air wavelengths except for the $\lambda2836.56$ line, which is vacuum.}
\label{f1}
\end{figure}

Throughout this paper air wavelengths are used for the optical lines and vacuum
wavelengths for the UV ones.
The transitions giving rise to the [\ion{Fe}{4}] lines discussed here
are identified in Figure~\ref{f1} (see also Table~\ref{t0}).

The deepest optical spectrum of \objectname{M42} has been published by
\citet{bal00}.
They find a weak feature at $\lambda6739.86$ which they consider as clearly
detected but unidentified. The line has an extinction-corrected
relative intensity of $I(\lambda6739.86)/I(\mbox{H}\beta)=3.7\times10^{-5}$.
The wavelength is very close to [\ion{Fe}{4}]~$\lambda6739.8$
($^2I_{11/2}\rightarrow{^4G}_{11/2}$), the brightest optical [\ion{Fe}{4}] line
that can be expected to form at the physical conditions prevailing in
\objectname{M42}.
This line was previously identified in the spectrum of
\objectname{NGC~7027} \citep{balu95}, one of the most dense and luminous
planetary nebulae, which has an extremely rich spectrum.
For the Orion observations the agreement in wavelength is very good,
especially when accounting for the difference of
$\sim2\mbox{~km~s}^{-1}$ due to the trend of velocity vs.\
ionization potential found by \citet{bal00}.
The measurements of [\ion{Fe}{4}]~$\lambda$2836.56 by \citet{rub97} and
[\ion{Fe}{4}]~$\lambda$6739.8 by \citet{bal00} were performed at different
positions in Orion, but both positions are at similar distances from the
ionizing star \objectname{$\theta^1$~Ori~C} (32\arcsec\ and 37\arcsec,
respectively), so that it might be instructive to compare their intensities.
The reddening-corrected intensities relative to H$\beta$ of both lines imply
$I(\lambda6739.8)/I(\lambda2836.56)=0.039$, whereas for typical Orion physical
conditions (see Table~\ref{t1} and \S3 below) the expected value of this ratio
is $\simeq0.022$.
Other optical [\ion{Fe}{4}] lines following $\lambda$6739.8 in intensity would
be $\lambda$4906.56 ($^4F_{9/2}\rightarrow{^4G}_{11/2}$), $\lambda$6734.4
($^2I_{13/2}\rightarrow{^4G}_{11/2}$), and $\lambda$6761.3
($^2I_{11/2}\rightarrow{^4G}_{9/2}$), but these
lines have expected intensities of about half the $\lambda$6739.8 intensity
and they are at the limit of line detection of \citet{bal00}.
We have then a good case for the detection of the first optical [\ion{Fe}{4}]
line in \objectname{M42}.

A literature search for other nebulae in which to perform an analysis of
[\ion{Fe}{4}] emission and the Fe abundance yielded four suitable objects:
\objectname{IC~4846}, \objectname{30~Doradus}, \objectname{SBS~0335$-$052}, and
\objectname{SMC~N88A}.
Available spectra of these objects include one [\ion{Fe}{4}] line, one or more
[\ion{Fe}{3}] lines, and the [\ion{O}{2}], [\ion{O}{3}], and [\ion{S}{2}] lines
needed to derive $T_{\rm e}$ values and electron density ($N_{\rm e}$) values,
the O abundance, and the degree of ionization $\mbox{O}^+/\mbox{O}$.
The optical [\ion{Fe}{4}] lines measured by \citet{est02} in
\objectname{NGC~2363} and \objectname{NGC~5471} could not be used because the
[\ion{O}{2}]~$\lambda3727$ lines were outside their observed wavelength range.
Other objects, like the symbiotic nova \objectname{RR~Telecopii}, where several
[\ion{Fe}{4}] lines have been detected \citep[see, for example,][]{mck97}, were
excluded because the high densities responsible for their rich spectra ($N_{\rm
e}\sim10^6$ cm$^{-3}$) prevent the application of the usual diagnostics.

\objectname{IC~4846} is a compact planetary nebula, where \citet*{hyu01} measure
an unidentified feature at $\lambda6739.14$.
The difference in wavelength with [\ion{Fe}{4}]~$\lambda6739.8$ is compatible
with the differences that \citeauthor{hyu01}\ find for other well identified lines.
Since [\ion{Fe}{4}]~$\lambda6739.8$ is the brightest optical [\ion{Fe}{4}] line
for the physical conditions in \objectname{IC~4846} (see Table~\ref{t1}),
this is a likely identification for the measured line.
The relative intensity of the line after the correction for extinction is
$I(\lambda6739.14)/I(\mbox{H}\beta)=3.1\times10^{-4}$.

The spectra of \objectname{30~Doradus} observed by \citet{ape03} shows a line at
$\lambda\sim6734$ that he identifies as [\ion{Cr}{4}]~$\lambda6733.9$ with a
possible contribution of \ion{C}{2}]~$\lambda6734$.
Since for the physical conditions of this object (see Table~\ref{t1})
[\ion{Fe}{4}]~$\lambda6734.4$ ($^2I_{13/2}\rightarrow{^4G}_{11/2}$) would be
the brightest optical [\ion{Fe}{4}] line, the measured feature could have a
contribution from this line and the intensity of the feature can be used to
obtain an upper limit to the Fe$^{3+}$ abundance.
However, a lower upper limit can be obtained by noting the absence of
[\ion{Fe}{4}]~$\lambda6739.8$, the optical line that should follow
[\ion{Fe}{4}]~$\lambda6734.4$ in intensity, and taking as an upper limit to its
intensity the intensity measured for a nearby weak line,
[\ion{Cr}{4}]~$\lambda6747.5$:
$I(\lambda6739.8)/I(\mbox{H}\beta)\le1.05\times10^{-4}$.
There is a feature at $\lambda5032.4$ identified by \citet{ape03} as partly due
to [\ion{Fe}{4}]~$\lambda5032.4$ ($^2F_{5/2}\rightarrow{^4G}_{7/2}$), but
this line should have a negligible intensity at low densities according to the
current atomic data and, therefore, the identification seems unlikely.

The blue compact dwarf galaxy \objectname{SBS~0335$-$052} has been observed by
\citet{izo01}, who tentatively identify a line measured at $\lambda4904$ as
[\ion{Fe}{4}]~$\lambda4906.56$.
Given the low spectral resolution, 8~\AA, and considering the physical
conditions in \objectname{SBS~0335$-$052} (see Table~\ref{t1}), the line is most
likely a blend of three [\ion{Fe}{4}] lines in the multiplet: $\lambda4899.97$
($^4F_{7/2}\rightarrow{^4G}_{9/2}$), $\lambda4903.07$
($^4F_{7/2}\rightarrow{^4G}_{7/2}$), and $\lambda4906.56$
($^4F_{9/2}\rightarrow{^4G}_{11/2}$).
Two of the unidentified features measured by \citet{izo01}, at $\lambda5235$ and
$\lambda7224$, could also be [\ion{Fe}{4}] lines: $\lambda5233.76$
($^2F_{7/2}\rightarrow{^4G}_{9/2}$), and $\lambda7222.8$
($^4F_{9/2}\rightarrow{^4D}_{7/2}$).
The intensities of these two lines relative to the blend at $\lambda4904$
are consistent with the expected values within a factor of 2 (see
Table~\ref{t0}).
Other [\ion{Fe}{4}] lines that could be present in the spectrum of
\objectname{SBS~0335$-$052} are likely to be blended with stronger lines.
Since the uncertainties assigned by \citet{izo01} to the intensities of the
weak lines are $\sim100\%$, no further assessment can be made on the reliability
of the relative line intensities predicted by the atomic data for [\ion{Fe}{4}].
The $\mbox{Fe}^{3+}$ abundance in this object will be derived from the strongest
feature, the [\ion{Fe}{4}] blend at $\lambda4904$, which \citet{izo01} consider
as clearly detected.
The extinction-corrected intensity of this feature relative to H$\beta$ is
$I(\lambda4904)/I(\mbox{H}\beta)=2.2\times10^{-3}$.

The UV and optical spectra of the Small Magellanic Cloud \ion{H}{2} region
\objectname{SMC~N88A} have been obtained by \citet{kurt99}, who measure
[\ion{Fe}{4}]~$\lambda$2836.56 at two positions in the nebula.
The values for the extinction-corrected line ratios are
$I(\lambda2837)/I(\mbox{H}\beta)=2.5\times10^{-2}$ and $2.4\times10^{-2}$
for the positions identified as bar and square~A both in their paper and here.

\section{RESULTS}

\subsection{Atomic Data}

The calculations for [\ion{Fe}{4}] used throughout this paper are based on a
33-level model-atom where all collisional and downward radiative
transitions are considered.
The collision strengths are those calculated by \citet{zha97}, the transition
probabilities those recommended by \citet{fro98} (and those from
\citealt{gar58} for those transitions not considered by \citeauthor{fro98}),
and the level energies have been taken from the
NIST database\footnote{Available at
\url{http://Physics.nist.gov/cgi-bin/AtData/main$_-$asd}.}.
The calculations for [\ion{Fe}{3}] are based on a
34-level model-atom that uses the collision strengths of \citet{zha96}
and the transition probabilities of \citet{qui96}.
The physical conditions and the abundances of the O ions
have been derived with the {\em nebular\/} package in IRAF\@.

\subsection{Errors and Uncertainties}

The abundances presented here are affected by the usual uncertainties related
to the method of calculation: (i) the assumption that the observed lines
originate in one or two emitting layers of constant $N_{\rm e}$ and
$T_{\rm e}$, (ii) the uncertainties in the atomic data used to derive physical
conditions and ionic abundances, (iii) errors in the line intensities, and (iv)
the uncertainties arising from the use of ICFs to
account for unseen stages of ionization.

The systematic uncertainties arising from any of these causes are very
difficult to estimate.
The errors presented here have been calculated by considering only the errors in
the line intensities, following the guidelines given by the different authors
for each object, but taking 5\% as the minimum error for any line intensity
relative to H$\beta$.
It should be kept in mind that the errors in the line intensity ratios are just
estimates, and that the criteria followed by the different authors may vary.
The errors for each calculated quantity have been derived by adding
quadratically the errors in the line intensity ratio used to derive the ionic
abundance and the errors arising from the uncertainties in both $T_{\rm e}$
and $N_{\rm e}$.
The last ones are especially important for the O$^+$ abundance and all ratios
involving this ion (e.g. $\mbox{O}^{+}/\mbox{O}^{++}$,
$\mbox{Fe}^{++}/\mbox{O}^{+}$).
Thus, the calculated errors can be used to assess the
sensitivity of each quantity to the adopted uncertainties.

\subsection{Physical Conditions and Ionic Abundances}

Several diagnostic lines were available for \objectname{30~Doradus},
\objectname{IC~4846} and \objectname{M42}.
For these objects the $T_{\rm e}$ obtained from the diagnostic [\ion{N}{2}] and
[\ion{O}{3}] lines have been used to derive the ionic abundances of the low and
high ionization species, respectively.
The values of the [\ion{N}{2}] and [\ion{O}{3}] $T_{\rm e}$ are listed in
Table~\ref{t1}.
The $N_{\rm e}$ values listed for these three objects have been derived from
the [\ion{S}{2}], [\ion{O}{2}], [\ion{Cl}{3}], and [\ion{Ar}{4}] diagnostic
lines.
Table~\ref{t1} shows the mean and standard deviation of the $N_{\rm e}$ values
obtained from the different line ratios.
For \objectname{SMC~N88A} and \objectname{SBS~0335$-$052} the values used for
$T_{\rm e}$ and $N_{\rm e}$ were those derived from the [\ion{O}{3}] and
[\ion{S}{2}] diagnostic lines.
The upper limit of $N_{\rm e}\mbox{[\ion{S}{2}]}$ in
\objectname{SMC-N88A}~bar was unconstrained with the errors found for the ratio
of [\ion{S}{2}] lines $I(\lambda6716)/I(\lambda6731)$. The upper limit of
$N_{\rm e}$ provided in Table~\ref{t1} for this object was obtained from the
constraints imposed on $N_{\rm e}$ by other line ratios.

Table~\ref{t0} shows the [\ion{Fe}{3}] and [\ion{Fe}{4}] lines used for the
determination of the $\mbox{Fe}^{++}$ and $\mbox{Fe}^{3+}$ abundances in the
different objects.
In \objectname{30~Doradus} and \objectname{M42}, 13 to 14 [\ion{Fe}{3}] lines
were considered to be unblended and available for the abundance determination.
The mean values and standard deviations of the calculated $\mbox{Fe}^{++}$
abundances are
$\mbox{Fe}^{++}/\mbox{H}^{+}=1.67\pm0.21\times10^{-7}$ for
\objectname{30~Doradus} and
$\mbox{Fe}^{++}/\mbox{H}^{+}=3.29\pm0.21\times10^{-7}$ for \objectname{M42}.
The agreement between the lines is extremely good, suggesting that the atomic
data used for $\mbox{Fe}^{++}$ are quite reliable \citep[see also][]{rod02}.
As for the other objects, three [\ion{Fe}{3}] lines could be used for
\objectname{IC~4846} and just one, [\ion{Fe}{3}]~$\lambda4658.1$, for
\objectname{SMC~N88A} and \objectname{SBS~0335$-$052}.
[The intensity given by \citealt{izo01} for [\ion{Fe}{3}]~$\lambda5270.4$ in
\objectname{SBS~0335$-$052} is clearly wrong, as can be seen by inspecting their
Fig.~1, but the weaker [\ion{Fe}{3}]~$\lambda4754.7$ line agrees to
within 10\% with the $\mbox{Fe}^{++}$ abundance implied by
[\ion{Fe}{3}]~$\lambda4658.1$ (see Table~\ref{t0}).]

The intensities measured by \citet{kurt99} for
[\ion{Fe}{4}]~$\lambda$2836.56 in \objectname{SMC~N88A} could have a
contribution from \ion{C}{2}~$\lambda\lambda2837.5$, 2838.4.
In \objectname{M42}, these recombination lines account for nearly 90\% of the
blend intensity \citep{rub97}.
The contribution of the \ion{C}{2} lines to the $\lambda$2837 feature in
\objectname{SMC~N88A} can be estimated using the recombination
coefficients of \citet{dav00} and the  C$^{++}$ abundances derived by
\citeauthor{kurt99}\ from \ion{C}{3}]~$\lambda1909$.
Taking into account that in \ion{H}{2} regions the C$^{++}$ abundances
implied by recombination lines are usually a factor $\sim2$ higher than those
derived from collisionally excited lines \citep[see, for example, Table~6
in][]{est02}, I estimate that the contribution of the \ion{C}{2} lines to the
intensity measured for [\ion{Fe}{4}]~$\lambda$2836.56 in \objectname{SMC~N88A}
is below 10\%.
This contribution will be neglected in what follows.

The large difference between the relative intensities of the [\ion{Fe}{4}] and
\ion{C}{2} lines in \objectname{M42} and \objectname{SMC~N88A} arises from the
lower metallicity of the latter (by a factor $\sim3.5$ in the O abundance, see
Table~\ref{t2}) and its associated higher $T_{\rm e}$.
The intensity of the C recombination lines decreases because of the lower
metallicity and higher $T_{\rm e}$, whereas the intensity of the forbidden line
increases because its emissivity shows an exponential dependence on
$T_{\rm e}$.
The same argument can be used to rule out any significant contamination of
either [\ion{Fe}{3}]~$\lambda4658.1$ or [\ion{Fe}{4}]~$\lambda4904$ by
\ion{O}{2}~$\lambda4661.63$ and \ion{O}{2}~$\lambda$4906.83 in
\objectname{SMC~N88A} and \objectname{SBS~0335$-$052}.
In \objectname{SBS~0335$-$052}, this conclusion is further confirmed by the
overall agreement shown by the other [\ion{Fe}{3}] and [\ion{Fe}{4}] lines (see
Table~\ref{t0}).

The final adopted values for the ionic abundances of $\mbox{O}^{+}$,
$\mbox{O}^{++}$, $\mbox{Fe}^{++}$, and $\mbox{Fe}^{3+}$ are listed in
Table~\ref{t1}.

\subsection{Total Abundances}

The ion Fe$^{+}$, whose ionization potential is 16.2~eV, can make some
contribution to the Fe abundance.
Although most of the [\ion{Fe}{2}] lines in \ion{H}{2} regions are affected by
fluorescence effects \citep[and references therein]{rod99a}, an
estimate of the Fe$^{+}$ contribution to the total Fe abundance can be
obtained from the intensity of [\ion{Fe}{2}]~$\lambda$8617, a line almost
insensitive to the effects of UV pumping.
This line is either undetected or out of the wavelength range measured for the
objects considered here.
An estimate of the Fe$^{+}$ abundance in \objectname{M42} was obtained by
averaging the values of $\mbox{Fe}^{+}/\mbox{H}^{+}$ found in \citet{rod02} for
three regions in \objectname{M42} that are 27\arcsec\ south of the position
observed by \citet{bal00}, but at a similar distance from
\objectname{$\theta^1$~Ori~C}.
The resulting value, $\mbox{Fe}^{+}/\mbox{H}^{+}=5.4\pm2.7\times10^{-8}$, is
just $\sim15$\% of the abundance derived for Fe$^{++}$.
This estimate of the Fe$^{+}$ abundance in \objectname{M42} has been used in the
analysis described below, which leads to a discrepancy between the expected and
measured values of the $\mbox{Fe}^{3+}$ abundance of a factor $4.4$.
If $\mbox{Fe}^{+}$ had been neglected in this analysis, the discrepancy would be
lower, a factor $3.8$.
Since the other objects have a higher degree of ionization than
\objectname{M42}, the contribution of $\mbox{Fe}^{+}$ to their Fe abundance will
be neglected.
The effect of neglecting the $\mbox{Fe}^{+}$ abundance in the results for these
higher ionization objects is likely to be even lower than for \objectname{M42},
especially for \objectname{IC~4846} and \objectname{SMC~N88A}, where no
[\ion{Fe}{2}] lines are detected.
In any case, if the concentration of $\mbox{Fe}^{+}$ in any of these objects
were not negligible, the discrepancy between the expected and calculated
values of $\mbox{Fe}^{3+}$ would be higher than the values given below.

On the other hand, \objectname{IC~4846} and \objectname{SBS~0335$-$052} show
\ion{He}{2} emission in their spectra.
Since He$^+$, O$^{++}$ and Fe$^{3+}$ have similar ionization potentials
(54.4, 54.9, and 54.8~eV, respectively), the presence of
He$^{++}$ suggests that O$^{3+}$ and Fe$^{4+}$ might also be present.
In fact, \objectname{SBS~0335$-$052} shows emission in [\ion{Fe}{5}]~$\lambda$4227
and, possibly, in some [\ion{Fe}{6}] and [\ion{Fe}{7}] lines whose origin is not
clear \citep{izo01}.
The [\ion{Fe}{5}]~$\lambda$4227 line cannot be used at the moment to derive the
Fe$^{4+}$ abundance, since the required atomic data are not available.
However, the amount of He$^{++}$ is low in \objectname{SBS~0335$-$052}
($\mbox{He}^{++}/\mbox{He}^{+}\simeq0.025$) and lower in \objectname{IC~4846}
($\mbox{He}^{++}/\mbox{He}^{+}\simeq0.0053$), suggesting that the concentrations
of O$^{3+}$ and Fe$^{4+}$ are negligible.
Therefore, it will be assumed that $\mbox{O}/\mbox{H}=
\mbox{O}^+/\mbox{H}^++\mbox{O}^{++}/\mbox{H}^+$ and $\mbox{Fe}/\mbox{H}=
\mbox{Fe}^{++}/\mbox{H}^++\mbox{Fe}^{3+}/\mbox{H}^+$ (except for
\objectname{M42} where $\mbox{Fe}/\mbox{H}=
\mbox{Fe}^{+}/\mbox{H}^++\mbox{Fe}^{++}/\mbox{H}^++\mbox{Fe}^{3+}/\mbox{H}^+$,
with $\mbox{Fe}^{+}/\mbox{H}^{+}=5.4\pm2.7\times10^{-8}$).

Table~\ref{t2} shows the values of the total abundances for all the objects.
Two values are given for the Fe abundance. The first one has been derived from
[\ion{Fe}{3}] and an ICF, $\mbox{Fe}/\mbox{H}=
1.1\,[(\mbox{Fe}^++\mbox{Fe}^{++})/\mbox{O}^+]\,\mbox{O}/\mbox{H}$.
The ICF is based on the grids of ionization models
of \citet{sta90} and \citet{gru92}, and is further discussed in $\S4$ below.
The second value for the Fe abundance is just the sum of the derived ionic
abundances, $\mbox{Fe}/\mbox{H}=
\mbox{Fe}^+/\mbox{H}^+ +\mbox{Fe}^{++}/\mbox{H}^++\mbox{Fe}^{3+}/\mbox{H}^+$.
The Fe abundances based on the sum of the ionic abundances can be seen to be
systematically lower, by factors in the range 2.6--5.9, than those implied by
the $\mbox{Fe}^{++}$ abundance and an ICF.
The expected values of the $\mbox{Fe}^{3+}$ abundance,
$\mbox{Fe}^{3+}_{\rm exp}$, can be obtained from the relation
$\mbox{Fe}^+/\mbox{H}^+ + \mbox{Fe}^{++}/\mbox{H}^+
+\mbox{Fe}^{3+}_{\rm exp}/\mbox{H}^+=
1.1\,[(\mbox{Fe}^++\mbox{Fe}^{++})/\mbox{O}^+]\,\mbox{O}/\mbox{H}$.
The values found for $\mbox{Fe}^{3+}_{\rm exp}/\mbox{Fe}^{3+}$ are shown in
Table~\ref{t2}:
the derived $\mbox{Fe}^{3+}$ abundances are lower than expected by
factors $\ge2.7$
(\objectname{30~Doradus}), $3.2$ (\objectname{IC~4846}), $4.4$
(\objectname{M42}), $5.5$ (\objectname{SMC~N88A}~bar),
7.5 (\objectname{SMC~N88A} square~A), and 6.1 (\objectname{SBS~0335$-$052}).

\section{IONIZATION CORRECTION FACTORS}

The above comparison between the expected and calculated values of
$\mbox{Fe}^{3+}/\mbox{H}^+$ relies on the ICF applied
to the $\mbox{Fe}^{++}$ abundance  in order to obtain the total Fe abundance.
Since [\ion{Fe}{3}] lines are the Fe lines most easily detected in \ion{H}{2}
regions, this ICF is a key parameter in the
determination of the Fe abundance in these nebulae.
The O ions are probably the best choice for defining ICFs for the Fe ions.
First, because both $\mbox{O}^{+}$ and $\mbox{O}^{++}$ can be easily measured
from strong optical lines.
Second, because the ionization potentials for the Fe and O ions are not too far
apart; 16.2, 30.6, and 54.8 eV for $\mbox{Fe}^{+}$, $\mbox{Fe}^{++}$, and
$\mbox{Fe}^{3+}$; 13.6, 35.1, and 54.9 eV for $\mbox{O}^{0}$,
$\mbox{O}^{+}$ and $\mbox{O}^{++}$.
The relations between the Fe and O ions and the $\mbox{Fe}/\mbox{O}$ abundance
ratio can be expressed in the following way:

\begin{equation}
\frac{\mbox{Fe}}{\mbox{O}}=\Bigg[\frac{x(\mbox{O}^+)}{x(\mbox{Fe}^{++})}\Bigg]\,
	\frac{\mbox{Fe}^{++}}{\mbox{O}^+},
\end{equation}

\begin{equation}
\frac{\mbox{Fe}}{\mbox{O}}=
	\Bigg[\frac{x(\mbox{O}^+)}{x(\mbox{Fe}^{+}+\mbox{Fe}^{++})}\Bigg]\,
	\frac{\mbox{Fe}^{+}+\mbox{Fe}^{++}}{\mbox{O}^+},
\end{equation}

\begin{equation}
\frac{\mbox{Fe}}{\mbox{O}}=\Bigg[\frac{x(\mbox{O}^{++})}{x(\mbox{Fe}^{3+})}\Bigg]\,
	\frac{\mbox{Fe}^{3+}}{\mbox{O}^{++}},
\end{equation}

\noindent
where $x(X)$ is the ionization fraction for the corresponding ion, and the
quantities in square brackets are ICFs that will be
constant if the ionization fractions of Fe and O vary in similar ways.

The values of the ICFs $[x(\mbox{O}^+)/x(\mbox{Fe}^{++})]$,
$[x(\mbox{O}^+)/x(\mbox{Fe}^{+}+\mbox{Fe}^{++})]$, and
$[x(\mbox{O}^{++})/x(\mbox{Fe}^{3+})]$ are shown in Figure~\ref{f2} as a
function of $\mbox{O}^{+}/\mbox{O}^{++}$ for the two series of models
calculated by \citet{gru92} and \citet{sta90}.
The ionization models from \citet{gru92} have metallicities Z$_\odot$,
Z$_\odot/3$, Z$_\odot/10$, and Z$_\odot/100$, electron densities of
10, 100, and 1000~cm$^{-3}$, and are ionized by a single star with effective
temperature $T_{\rm eff}=30900$ or 50000~K.
The models from \citet{sta90} considered in Figure~\ref{f2} are those ionized
by one star for Z$_\odot$ and Z$_\odot/2$, and those ionized by 1, $10^2$ or
$10^4$ stars for Z$_\odot/5$, Z$_\odot/10$, and Z$_\odot/50$. These models have
$N_{\rm e}$ of 10 and 1000~cm$^{-3}$, and $T_{\rm eff}=32500\mbox{--}55000$~K.
The main difference between the two series is that the results of \citet{gru92}
are presented for several lines of sight across each model.

The results from the two series of models shown in Figure~\ref{f2} can be compared with
those of three individual models for \objectname{M42}:
$\log[x(\mbox{O}^+)/x(\mbox{Fe}^{++})]=0.26$
\citep{bal91}, $\log[x(\mbox{O}^+)/x(\mbox{Fe}^{++})]=0.15$
\citep{rub91a,rub91b},
and $\log[x(\mbox{O}^+)/x(\mbox{Fe}^{++})]=0.04$ \citep{bau98}.
The last result has been calculated with the most recent values for the
photoionization cross-sections and recombination-rate coefficients for the Fe
ions \citep{nah96a,nah96b}, and is similar to the ICF used here.

Figure~\ref{f2} also shows the results obtained here for the five studied
objects (the filled squares for \objectname{IC~4846}, \objectname{M42},
\objectname{SMC~N88A}, and \objectname{SBS~0335$-$052};
the lower and upper limits for \objectname{30~Doradus})
by assuming that $\mbox{Fe}/\mbox{H}=
\mbox{Fe}^{+}/\mbox{H}^+ +\mbox{Fe}^{++}/\mbox{H}^++\mbox{Fe}^{3+}/\mbox{H}^+$.

Several comments can be made about Figure~\ref{f2}:
\begin{enumerate}
\item{The results for those models of \citet{gru92} with solar
metallicity deviate from the relation followed by the models with lower
metallicities. This suggests that the ICFs might be
dependent on metallicity, but the models of \citet{sta90} and the results of
\citet{gru92} for subsolar metallicities do not show such dependence.}

\item{Although the results for each series of models are consistent with roughly
constant values for $[x(\mbox{O}^+)/x(\mbox{Fe}^{++})]$ and 
$[x(\mbox{O}^+)/x(\mbox{Fe}^{+}+\mbox{Fe}^{++})]$, the values are different
for each series.}

\item{The values of $[x(\mbox{O}^+)/x(\mbox{Fe}^{++})]$ are substantially
higher than those of $[x(\mbox{O}^+)/x(\mbox{Fe}^{+}+\mbox{Fe}^{++})]$. This
is due to the fact that the models predict a significant concentration of
Fe$^+$, although with great scatter: $0.05<(\mbox{Fe}^+/\mbox{Fe}^{++})<0.7$.
However, the values found in \citet{rod02} for Galactic \ion{H}{2} regions with
$\log(\mbox{O}^{+}/\mbox{O}^{++})<0$ for $\mbox{Fe}^+/\mbox{Fe}^{++}$ are all
lower than 0.3 and mostly around 0.1.
The concentrations of ions with low ionization potential like Fe$^+$ are very
model-dependent and, therefore, difficult to estimate with reliability.
Furthermore, as suggested by the referee, this difference between the expected
and calculated concentrations of Fe$^+$ can be due to the presence of
significant amounts of Fe$^+$ in neutral zones, where the $T_{\rm e}$ is too
low to produce the optical [\ion{Fe}{2}] lines.
Therefore, I consider it more reliable to use the ICF implied by
$[x(\mbox{O}^+)/x(\mbox{Fe}^{+}+\mbox{Fe}^{++})]$, neglecting the contribution
of $\mbox{Fe}^{+}$ for the high-ionization objects.
As discussed above, this should be a good approximation for \objectname{IC~4846}
and \objectname{SMC~N88A}, where no [\ion{Fe}{2}] lines are
detected.
If the contribution of $\mbox{Fe}^{+}$ were higher than assumed, the
discrepancy between the expected and calculated values of $\mbox{Fe}^{3+}$ would
be even higher than the values derived here.
The calculated results are in any case systematically below the expected ones.}

\item{According to both series of models,
$[x(\mbox{O}^{++})/x(\mbox{Fe}^{3+})]\simeq1$ for
$\log(\mbox{O}^{+}/\mbox{O}^{++})<0$, and therefore
$\mbox{Fe}^{3+}/\mbox{O}^{++}\simeq\mbox{Fe}/\mbox{O}$ to within 0.1~dex.
Since $\mbox{Fe}^{3+}$ and $\mbox{O}^{++}$ are formed at 30.6 and 35.1~eV, and
are ionized at 54.8 and 54.9~eV, for those conditions with
$x(\mbox{O}^{++})\ge0.9$, most of Fe should be present as $\mbox{Fe}^{3+}$.
Therefore, barring large errors in the atomic data used to derive the Fe
ionization balance, $\mbox{Fe}^{3+}/\mbox{O}^{++}\simeq\mbox{Fe}/\mbox{O}$
should be a very good approximation for high-ionization objects; whereas
for any degree of ionization it should hold that
$\mbox{Fe}^{3+}/\mbox{O}^{++}\ge\mbox{Fe}/\mbox{O}$.
However, as seen in Figure~\ref{f2}c, all the objects show
$\mbox{Fe}^{3+}/\mbox{O}^{++}<\mbox{Fe}/\mbox{O}$.}

\item{Although the error bars for \objectname{M42} and \objectname{IC~4846} are
almost consistent with the expected results for
$[x(\mbox{O}^+)/x(\mbox{Fe}^{+}+\mbox{Fe}^{++})]$ and
$[x(\mbox{O}^{++})/x(\mbox{Fe}^{3+})]$, the results for the other objects and
the fact that all the calculated results deviate in the same direction from the
expected ones confirm that there is a significant deviation.}

\item{If the atomic data used in the models to derive the ionization balance of
Fe were in error, the ionization fractions calculated for real objects could be
used to obtain an estimate of the actual ICFs.
The results in Figure~\ref{f2} can be interpreted in such a way. A weighted
least-squares fit to the data in Figure~\ref{f2}a leads to the following ICF,

\begin{equation}
\Bigg[\frac{x(\mbox{O}^+)}{x(\mbox{Fe}^{++})}\Bigg]\,=0.78\,
	\Bigg(\frac{\mbox{O}^+}{\mbox{O}^{++}}\Bigg)^{0.43},
\end{equation}

for $-1.5\le\log(\mbox{O}^+/\mbox{O}^{++})\le-0.5$, but the significance of the
fit is not large, and other alternatives such as a
constant $[x(\mbox{O}^+)/x(\mbox{Fe}^{++})]\simeq0.25$, cannot be
excluded.}

\end{enumerate}

In summary, there is a clear discrepancy between the calculated results and the
model predictions.
Even though the discrepancy might not be due to errors in the models (see \S5
below), there are some problems with the models that would be worth exploring
with further calculations.
The ICFs selected here are in any case those leading to
the lower discrepancies while at the same time being consistent with the model
results.

\section{DISCUSSION}

There are three possible explanations for the discrepancy in the
Fe abundances obtained from [\ion{Fe}{3}] and [\ion{Fe}{4}]:
(i) that the atomic data for [\ion{Fe}{4}] are unreliable; (ii) that the
concentrations for the Fe ions predicted by photoionization models are
greatly in error; and (iii) that there is some unknown mechanism
producing a gradient in the Fe abundance within the ionized gas, as suggested
by \citet{bau98}.
The high value of the discrepancy excludes other explanations, like errors in
the line intensities, errors in the calculated physical conditions, or errors
in the atomic data for [\ion{Fe}{3}], which seem reliable, as discussed in
$\S3.3$.
A contribution of fluorescence to [\ion{Fe}{3}] emission can also be ruled out
\citep{luc95,bau98}.

The [\ion{Fe}{4}] emissivities are almost insensitive to the values used for the
transition probabilities and depend mainly on the values of the collision
strengths.
Therefore, to explain the discrepancies between the expected and calculated
values for Fe$^{3+}$ ($\mbox{Fe}^{3+}_{\rm exp}/\mbox{Fe}^{3+}$ in
Table~\ref{t2}), the simplest solution would be to lower all the collision
strengths by a factor $\sim\mbox{Fe}^{3+}_{\rm exp}/\mbox{Fe}^{3+}$.
The values and errors given in Table~\ref{t2} for the discrepancies imply that
if the collision strengths were lower by a factor of $\sim5$, the Fe$^{3+}$
abundances would be consistent with the expected values.
On the other hand, there could be a difference between the discrepancies
obtained for \objectname{IC~4846} and \objectname{M42}
($\mbox{Fe}^{3+}_{\rm exp}/\mbox{Fe}^{3+}\sim3.8$) and those for the
other objects ($\mbox{Fe}^{3+}_{\rm exp}/\mbox{Fe}^{3+}\sim6$).
This difference does not seem to arise from the fact that the $T_{\rm e}$
implied by the [\ion{O}{3}] lines has been used to derive all the ionic
abundances in \objectname{SMC~N88A} and \objectname{SBS~0335$-$052}, whereas
$T_{\rm e}$[\ion{N}{2}] has been used in the other objects for deriving the
abundances of the low ionization ions.
An estimate of $T_{\rm e}$[\ion{N}{2}] in \objectname{SMC~N88A} and
\objectname{SBS~0335$-$052} can be obtained from $T_{\rm e}$[\ion{O}{3}] and the
relation derived by \citet*{cam86} from the models of \citet{sta82}.
If these estimates of $T_{\rm e}$[\ion{N}{2}] were used in the analysis, the
discrepancies for \objectname{SMC~N88A} and \objectname{SBS~0335$-$052} would
decrease, but by only a small amount to 5.1, 7.1 and 5.4 for
\objectname{SMC~N88A}~bar, \objectname{SMC~N88A}~sq.~A, and
\objectname{SBS~0335$-$052}, respectively.

The $\mbox{Fe}^{3+}$ abundance has been obtained for \objectname{IC~4846} and
\objectname{M42} from the intensity of the line [\ion{Fe}{4}]~$\lambda6739.8$,
whereas [\ion{Fe}{4}]~$\lambda$2836.56 and the [\ion{Fe}{4}] blend at
$\lambda4904$ have been used for \objectname{SMC~N88A} and
\objectname{SBS~0335$-$052} respectively.
The upper level of the transition [\ion{Fe}{4}]~$\lambda6739.8$, $^2I_{11/2}$,
is mainly populated by collisional excitations from the $^4G$ levels, which are
metastable.
The levels $^4F_{7/2}$ and $^4F_{9/2}$, giving rise to the blend at
$\lambda4904$, are populated by collisional excitations from the $^4G$ levels
and from the ground state.
The upper level of [\ion{Fe}{4}]~$\lambda$2836.56, $^4P_{5/2}$, is populated by
spontaneous emission from the $^4D$ term and by collisional excitations from
both the $^4G$ term and the ground state.
The different discrepancies obtained from [\ion{Fe}{4}]~$\lambda6739.8$, on the
one hand, and [\ion{Fe}{4}]~$\lambda$2836.56, $\lambda4904$, on the other, could
then be the effect of errors in the atomic data.
As an example, if the collision strengths involving only the Fe$^{3+}$ ground
state were lowered by a factor of 6.5, the expected and calculated values of
the Fe$^{3+}$ abundance would differ by less than $\sim50$\%.
However, the difference in the discrepancies could be due to other causes.
One possibility would be that the ICFs are highly
dependent on metallicity; another, that the difference between the Fe abundances
in the [\ion{Fe}{3}] and [\ion{Fe}{4}] emitting regions depends on the
metallicity or varies from object to object.

If the trend of increasing
$[x(\mbox{O}^+)/x(\mbox{Fe}^{++})]$ with $\mbox{O}^+/\mbox{O}^{++}$ suggested by
the calculated data in Figure~\ref{f2} were real, the Fe ionization fractions
predicted by models should be seriously questioned.
The value predicted by models for the relative concentrations of
$\mbox{Fe}^{++}$ and $\mbox{Fe}^{3+}$, $\mbox{Fe}^{++}/\mbox{Fe}^{3+}$, is
roughly proportional to the ratio between the total recombination coefficient
of Fe$^{3+}$ and the ionization cross-section for Fe$^{++}$ integrated over the
radiation field.
The latter ratio should then be higher by a factor $\sim5$ to explain the
discrepancy.
The recent calculations of the ionization and recombination cross-sections for
Fe$^{++}$/Fe$^{3+}$ \citep{nah96a,nah96b} are significantly different from the
previous data.
The new value for the total recombination coefficient for $\mbox{Fe}^{3+}$
\citep{nah96b} is a factor of 1.5 higher at $T_{\rm e}\sim10^4$~K than the
previous value by \citet*{woo81}.
On the other hand, the old values for the ionization cross-section of
$\mbox{Fe}^{++}$ \citep{rei79} were calculated for energies above 35~eV and,
when extrapolated to lower energies, lead to values which are higher than those
calculated by \citet{nah96a} by a factor of 5 near the ionization threshold.
However, according to \citet{bau98}, this overestimation compensates in part for
the contribution of the many resonant structures found by \citep{nah96a} at low
energies.
Therefore, the new data finally lead to similar values for the ICFs---at least
for the Orion model of \citet{bau98}, as commented in \S4 above.
More extensive calculations exploring the effect of the new cross-sections for
different degrees of ionization might be valuable.
The effect of charge-exchange reactions, whose rates are highly uncertain
\citep{kin96}, should also be explored.
Such calculations will be the subject of future work.

An error in the calculations of the Fe ionization balance would prove
to be the simplest explanation for the trend in Figure~\ref{f2}.
Some mechanism leading to the preferential destruction of dust grains in the
low ionization zones could also explain such a trend, but this explanation
seems rather ad hoc and less likely.

The accurate measurement of the relative intensities of several [\ion{Fe}{4}]
lines in various objects where the physical conditions can also be measured with
reasonable accuracy would help to distinguish between all these possibilities.
These measurements could be attempted in low metallicity \ion{H}{2} regions.
The high $T_{\rm e}$ values prevailing in these objects boost the intensities
of forbidden lines while the low metallicity reduces the possible contamination
with permitted lines, an important issue when trying to measure very weak
lines.

Figure~\ref{f3} shows the values of the $\mbox{Fe}/\mbox{O}$ abundance ratio
obtained from the Fe$^{++}$ abundance and an ICF (Figs.~\ref{f3}a and
\ref{f3}b) and from the Fe$^{++}$ and Fe$^{3+}$ abundances (Figs.~\ref{f3}c and
\ref{f3}d).
The results are plotted as a function of the O abundance (Figs.~\ref{f3}a and
\ref{f3}c) and of the ionization degree $\mbox{O}^+/\mbox{O}^{++}$
(Figs.~\ref{f3}b and \ref{f3}d).
The solar $\mbox{Fe}/\mbox{O}$ and $\mbox{O}/\mbox{H}$ abundance ratios are
shown in Figure~\ref{f3}a as a dotted circle \citep{hol01}.
The real value of the $\mbox{Fe}/\mbox{O}$ abundance in the gas of a given
object will be the result of two factors: the intrinsic value of
$\mbox{Fe}/\mbox{O}$ (in gas and dust) and the amount of Fe depleted in dust
grains.
The intrinsic value of $\mbox{Fe}/\mbox{O}$ in a given object depends on the
previous star formation history, but is expected to show less variation from
object to object than either $\mbox{Fe}/\mbox{H}$ or $\mbox{O}/\mbox{H}$.
In stars of our Galaxy, $\mbox{Fe}/\mbox{O}$ increases with metallicity and is
0.2 to 0.3~dex below solar when $\mbox{O}/\mbox{H}$ is around 1~dex below solar
\citep[see, for example,][]{nis02}.
Abundance analyses of stars in the Magellanic clouds show the same increment but
displaced by about $0.2$~dex towards higher values of $\mbox{Fe}/\mbox{O}$ at a
given metallicity \citep[][and references therein]{kor00,smi02}.
The intrinsic value of $\mbox{Fe}/\mbox{O}$ in the interstellar medium of the
Magellanic Clouds (that is, for \objectname{SMC~N88A} and
\objectname{30~Doradus}) might then be solar or up to $0.2$~dex above solar.
The intrinsic value of $\mbox{Fe}/\mbox{O}$ in \objectname{SBS~0335$-$052} is not
known.
Chemical evolution models for another low metallicity dwarf galaxy,
\objectname{I~Zw~18}, predict values for $\mbox{Fe}/\mbox{O}$ ranging from
$0.1$~dex above solar to about $0.7$~dex below solar \citep[see, for
example,][]{rec02}.
This wide range of possible values arises from the uncertainties in both the
star formation history and the iron yields due to massive stars, and makes it
impossible to draw a conclusion on the most likely value for
$\mbox{Fe}/\mbox{O}$ in \objectname{SBS~0335$-$052}.
On the other hand, good constraints on these two issues could be obtained from
the real value of the $\mbox{Fe}/\mbox{O}$ abundance ratio in
\objectname{SBS~0335$-$052} and other low metallicity blue compact galaxies.
A low amount of dust within the ionized gas of these low metallicity objects
can be inferred from the low or negligible extinction measured for them.
Therefore, the higher value of $\mbox{Fe}/\mbox{O}$ derived for
\objectname{SBS~0335$-$052} (the one obtained with [\ion{Fe}{3}] emission and an
ICF) favors a near solar value for the intrinsic
$\mbox{Fe}/\mbox{O}$ in the galaxy, whereas the lower $\mbox{Fe}/\mbox{O}$
implied by [\ion{Fe}{4}] emission favors an intrinsic $\mbox{Fe}/\mbox{O}$
about $0.7$~dex below solar.

Figures~\ref{f3}b and \ref{f3}d suggest an explanation for the different depletion
factors of the objects in the sample.
If \objectname{IC~4846} (the only planetary nebula in the sample) is excluded,
the values of $\mbox{Fe}/\mbox{O}$ implied by both procedures increase with the
degree of ionization.
The same behavior was found in \citet{rod96,rod02} for Galactic \ion{H}{2}
regions with near solar metallicity and it was interpreted as due to the release
of Fe atoms from dust grains by the action of energetic photons.
The same process may be responsible for the low Fe depletion factors in
\objectname{SMC~N88A} and \objectname{SBS~0335$-$052}, but the amount of dust
destruction and the slope of its dependence on the number of energetic photons
depend strongly on which are the real values of $\mbox{Fe}/\mbox{O}$.
Thus, the solution to the discrepancy found for [\ion{Fe}{4}] emission, has
implications for both the chemical evolution of low metallicity galaxies and the
evolution of dust in ionized nebulae.

\section{CONCLUSIONS}

The line [\ion{Fe}{4}]~$\lambda6739.8$ has been identified in published spectra
of \objectname{M42} and \objectname{IC~4846}.
Upper limits to the intensity of this line and of [\ion{Fe}{4}]~$\lambda6734.4$
have been established for \objectname{30~Doradus}.
The tentative identification by \citet{izo01} of a line at $\lambda\sim4904$ in
the spectra of \objectname{SBS~0335$-$052} as an [\ion{Fe}{4}] feature has been
confirmed.
These data along with the measurement by \citet{kurt99} of
[\ion{Fe}{4}]~$\lambda$2836.56 in two positions of \objectname{SMC~N88A} have
been used to perform an analysis of [\ion{Fe}{4}] emission in the five
aforementioned nebulae.
The Fe abundances obtained from [\ion{Fe}{3}] lines and an ICF derived from
ionization models, $\mbox{Fe}/\mbox{H}=
1.1\,[(\mbox{Fe}^++\mbox{Fe}^{++})/\mbox{O}^+]\,\mbox{O}/\mbox{H}$, have been
compared with those implied by the sum of the relevant ionic states,
$\mbox{Fe}/\mbox{H}=
\mbox{Fe}^+/\mbox{H}^++\mbox{Fe}^{++}/\mbox{H}^++\mbox{Fe}^{3+}/\mbox{H}^+$.
The $\mbox{Fe}/\mbox{H}$ abundance ratios obtained from the first method are
higher than those derived from the second method by factors in the range
2.6--5.9.
This result confirms the discrepancy previously found by \citet{rub97} in
\objectname{M42} between the Fe abundance implied by [\ion{Fe}{2}] and
[\ion{Fe}{3}] lines, and that implied by [\ion{Fe}{4}]~$\lambda2836.56$ .

\begin{figure}
\plotone{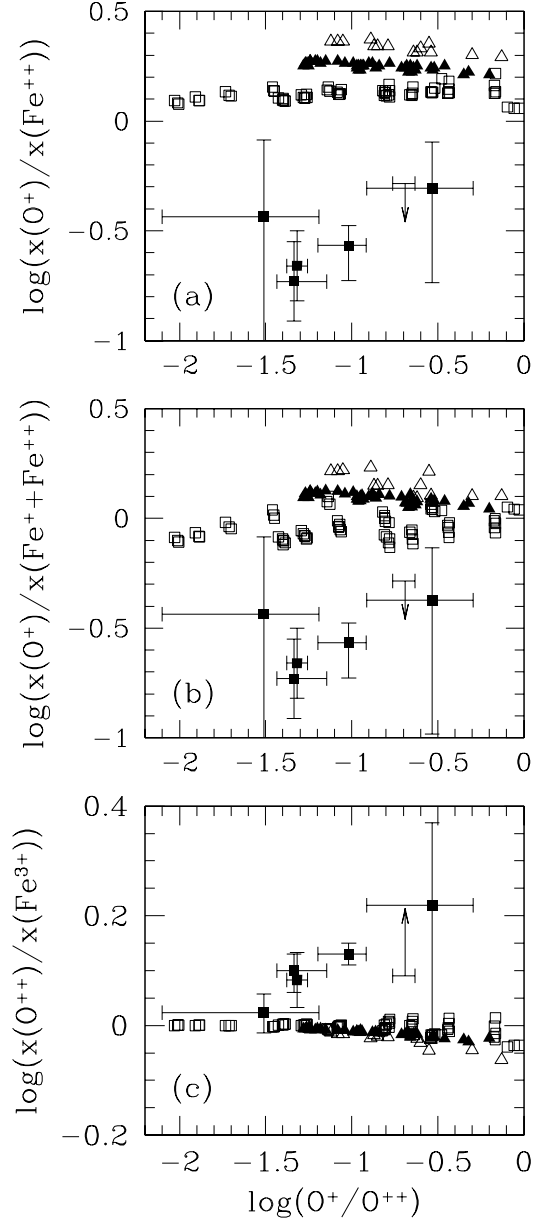}
\caption{Values of different ratios between the ionization fractions of the
Fe and O ions (the ICFs in eqs.\ [1], [2], and [3]) presented as a function of
the degree of ionization given by $\mbox{O}^+/\mbox{O}^{++}$.
The values calculated for IC~4846, M42, SMC~N88A, and SBS~0335$-$052 are shown as
filled squares; upper and lower limits are given for 30~Doradus, with the sizes
of the arrows indicating a decrease by a factor of 2 in the Fe$^{3+}$ abundance.
The Fe$^+$ abundance has been considered negligible for all objects except M42.
The empty squares are the results of the ionization models of \citet{sta90}.
The triangles show the results of the models of \citet{gru92}; empty triangles
have been used for those models with solar metallicity, which deviate somewhat
from the values for lower metallicities (the filled triangles).
More information on the selected models is given in the text.}
\label{f2}
\end{figure}

The $\mbox{Fe}^{3+}$ abundance is systematically lower than expected
for the five objects by factors from 3.2 to 7.5.
The uncertainties in the derived discrepancy factors are too high to reach a
definitive conclusion, but the present analysis offers two hints as to the
possible explanation:
\begin{enumerate}
\item{The discrepancies obtained with [\ion{Fe}{4}]~$\lambda6739.8$, on the one
hand, and [\ion{Fe}{4}]~$\lambda2836.56$ and the [\ion{Fe}{4}] blend at
$\lambda4904$, on the other, might be different (see the values of
$\mbox{Fe}^{3+}_{\rm exp}/\mbox{Fe}^{3+}$ in Table~\ref{t2}).
The measurement of these lines in a single object would help to establish this
issue.
If confirmed, this result would imply that the collision strengths for
$\mbox{Fe}^{3+}$ are unreliable.}
\item{The values of $[x(\mbox{O}^+)/x(\mbox{Fe}^{++})]$ derived for the objects
in the sample might show a trend with the degree of ionization given by
$\mbox{O}^+/\mbox{O}^{++}$ (see Fig.~\ref{f2}).
Since the ionization models predict a constant value for this ICF,
$[x(\mbox{O}^+)/x(\mbox{Fe}^{++})]\simeq1.1$, a deviation from
this constant value that depends on the degree of ionization would suggest that
the Fe ionization fractions predicted by models are seriously in error.
The measurement of [\ion{Fe}{4}] lines in more objects would help to establish
the reality of this trend.}
\end{enumerate}

Other explanations, like the existence of some kind
of gradient in the Fe abundance within the ionized gas, cannot be ruled out at
the moment.

\begin{figure*}
\plotone{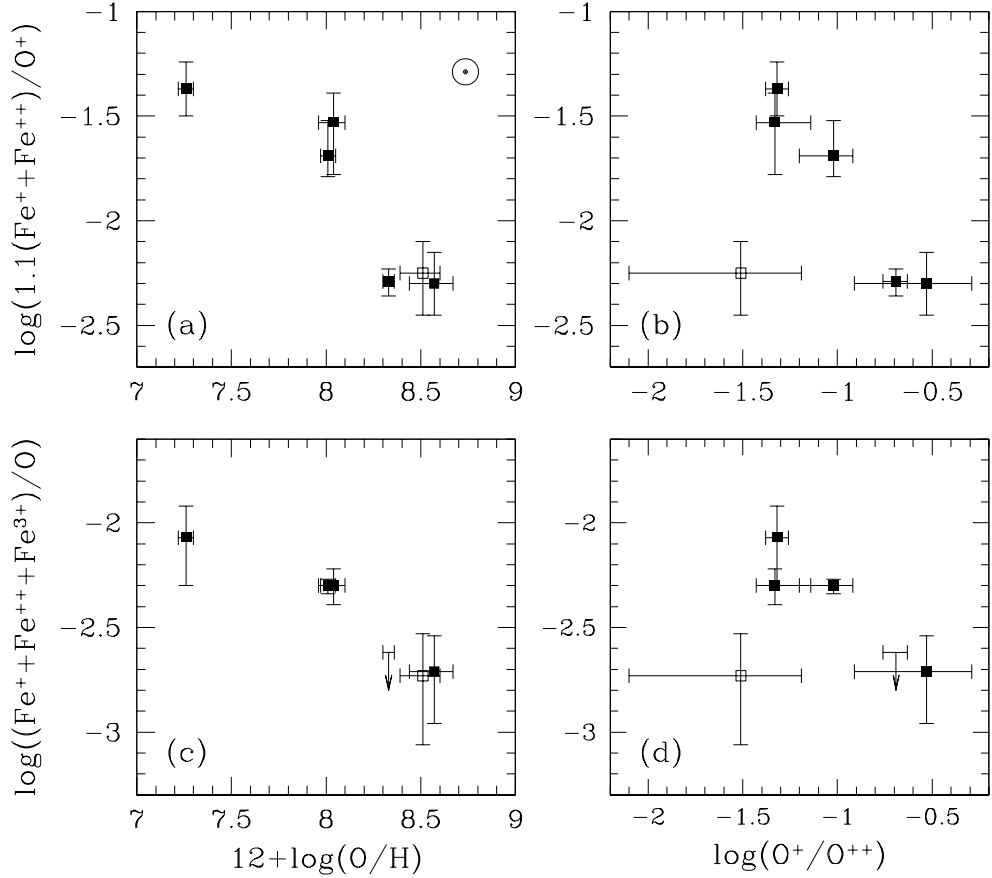}
\caption{$\mbox{Fe}/\mbox{O}$ abundance ratio as a function of the O abundance
and the degree of ionization for M42, 30~Doradus, IC~4846, two positions in
SMC~N88A, and SBS~0335$-$052. The values for the planetary nebula IC~4846 are
represented by an empty square. In panels ({\sl a\/}) and ({\sl b\/}) it has
been assumed that
$\mbox{Fe}/\mbox{O}=1.1\,(\mbox{Fe}^++\mbox{Fe}^{++})/\mbox{O}^+$; panels ({\sl
c\/}) and ({\sl d\/}) show the values obtained from
$\mbox{Fe}/\mbox{O}=(\mbox{Fe}^++\mbox{Fe}^{++}+\mbox{Fe}^{3+})/\mbox{O}$. The
contribution of $\mbox{Fe}^+$ has been considered negligible for all objects
except M42. The dotted circle in panel ({\sl a\/}) shows the solar abundances
\citep{hol01}; the size of this symbol represents approximately the associated
uncertainties.}
\label{f3}
\end{figure*}

The values of $\mbox{Fe}/\mbox{O}$ implied by both methods decrease with
metallicity, as shown in Figure~\ref{f3}.
This trend, which should be confirmed for other low metallicity objects,
probably reflects an increase of the Fe depletion factors in the different
objects as their metallicity increases.
The increment of Fe atoms in the gas of low metallicity \ion{H}{2} regions could
be due to the effect of the harder radiation fields typically found in these
objects.
This is suggested by the fact that if the planetary nebula \objectname{IC~4846}
is excluded, the $\mbox{Fe}/\mbox{O}$ abundance ratios follow and extend to
higher degrees of ionization the correlation with the degree of ionization
previously found in \citet{rod96,rod02} for Galactic \ion{H}{2} regions in the
solar neighborhood.
The deviation of \objectname{IC~4846} from the relationship could be due to the
large uncertainties in the abundances derived for this object or to the specific
origin and characteristics of dust grains in planetary nebulae.
Although the values of $\mbox{Fe}/\mbox{O}$ for the other objects follow the
correlation with the degree of ionization
independently of whether [\ion{Fe}{4}] emission is considered or not,
the shape of the correlation depends on which method is used in the abundance
determination.
Furthermore, the $\mbox{Fe}/\mbox{O}$ abundance ratio in the low metallicity
galaxy \objectname{SBS~0335$-$052}, which has important implications for our
understanding of chemical evolution, remains uncertain by a factor of 5.
All these implications emphasize the need for a correct understanding of the
reasons behind the [\ion{Fe}{4}] discrepancy.

\acknowledgments
I thank R.~Manso Sainz and R.~H.~Rubin for reading the paper and providing many
useful comments, improvements and corrections.
I also acknowledge discussions with L.~Binette and the valuable comments
of an anonymous referee.
This work has made use of NASA's Astrophysics Data System Abstract Service.
The project was supported by the Mexican CONACYT project J37680-E.

\clearpage

\clearpage

\begin{deluxetable}{lcccccc}
\tabletypesize{\footnotesize}
\tablecolumns{7}
\tablecaption{[\ion{Fe}{3}] and [\ion{Fe}{4}] Lines and the Abundances of
Fe$^{++}$ and Fe$^{3+}$ \label{t0}}
\tablewidth{0pt}
\tablehead{
\colhead{} & \multicolumn{6}{c}{$\mbox{Fe}^{++}/\mbox{H}^{+}$} \\[0.1cm]
\cline{2-7} \\
\colhead{[\ion{Fe}{3}] Line} & \colhead{30~Doradus} & \colhead{IC~4846} &
\colhead{M42} & \colhead{N88A bar}  & \colhead{N88A sq.~A} &
\colhead{SBS~0335$-$052}
}
\startdata
$\lambda$4008 ($^3G_4\rightarrow{^5D}_4$) & 1.89E$-7$
	& \nodata & 3.08E$-7$    & \nodata & \nodata & \nodata \\
$\lambda$4080 ($^3G_4\rightarrow{^5D}_3$) & \nodata
	& \nodata & 3.44E$-7$    & \nodata & \nodata & \nodata \\
$\lambda$4607 ($^3F_3\rightarrow{^5D}_4$) & 1.64E$-7$
	& \nodata & \nodata & \nodata & \nodata & \nodata \\
$\lambda$4658 ($^3F_4\rightarrow{^5D}_4$) & 1.59E$-7$
	& 3.03E$-8$ & 3.13E$-7$    & 1.69E$-7$ & 1.31E$-7$ & 3.24E$-8$ \\
$\lambda$4667 ($^3F_2\rightarrow{^5D}_3$) & \nodata
	& \nodata & 3.82E$-7$    & \nodata & \nodata & \nodata \\
$\lambda$4701 ($^3F_3\rightarrow{^5D}_3$) & 1.60E$-7$
	& \nodata & 3.15E$-7$    & \nodata & \nodata & \nodata \\
$\lambda$4734 ($^3F_2\rightarrow{^5D}_2$) & 1.75E$-7$
	& \nodata & 3.45E$-7$    & \nodata & \nodata & \nodata \\
$\lambda$4755 ($^3F_4\rightarrow{^5D}_3$) & 1.60E$-7$
	& \nodata & 3.21E$-7$ &\nodata & \nodata & 4.44E$-8$\tablenotemark{a}\\
$\lambda$4769 ($^3F_3\rightarrow{^5D}_2$) & 2.04E$-7$
	& \nodata & 3.17E$-7$    & \nodata & \nodata & \nodata \\
$\lambda$4778 ($^3F_2\rightarrow{^5D}_1$) & 1.41E$-7$
	& \nodata & 3.44E$-7$    & \nodata & \nodata & \nodata \\
$\lambda$4881 ($^3H_4\rightarrow{^5D}_4$) & 2.03E$-7$
	& 3.79E$-8$ & 3.21E$-7$    & \nodata & \nodata & \nodata \\
$\lambda$4986 ($^3H_6\rightarrow{^5D}_4$) & 1.40E$-7$
	& \nodata & \nodata & \nodata & \nodata & \nodata \\
$\lambda$5011 ($^3P_1\rightarrow{^5D}_2$) & 1.58E$-7$
	& \nodata & 3.10E$-7$    & \nodata & \nodata & \nodata \\
$\lambda$5084 ($^3P_1\rightarrow{^5D}_0$) & \nodata
	& \nodata & 3.44E$-7$    & \nodata & \nodata & \nodata \\
$\lambda$5270 ($^3P_2\rightarrow{^5D}_3$) & 1.69E$-7$
 	& 7.36E$-8$ & 3.07E$-7$    & \nodata & \nodata & \nodata \\
$\lambda$5412 ($^3P_2\rightarrow{^5D}_1$) & 1.48E$-7$
 	& \nodata & 3.38E$-7$    & \nodata & \nodata & \nodata \\[0.1cm]
\cline{1-7} \\
\colhead{} & \multicolumn{6}{c}{$\mbox{Fe}^{3+}/\mbox{H}^{+}$} \\[0.1cm]
\cline{2-7} \\
\colhead{[\ion{Fe}{4}] Line} & \colhead{30~Doradus} & \colhead{IC~4846} &
\colhead{M42} & \colhead{N88A bar}  & \colhead{N88A sq.~A} &
\colhead{SBS~0335$-$052} \\[0.1cm]
\cline{1-7} \\
$\lambda$2837 ($^4P_{5/2}\rightarrow{^6S}_{5/2}$) &
	\nodata & \nodata & \nodata & 3.51E$-7$ & 4.17E$-7$ & \nodata \\
$\lambda$4900 $(^4F_{7/2}\rightarrow{^4G}_{9/2})+$ & & & & & & \\
 $\lambda$4903 $(^4F_{7/2}\rightarrow{^4G}_{7/2})+$ & & & & & & \\
 $\lambda$4907 $(^4F_{9/2}\rightarrow{^4G}_{11/2})$ &
 	\nodata & \nodata & \nodata & \nodata & \nodata & 1.21E$-7$ \\
$\lambda$5234 ($^2F_{7/2}\rightarrow{^4G}_{9/2}$) &
	\nodata & \nodata & \nodata
& \nodata & \nodata & 1.39E$-7$\tablenotemark{a} \\
$\lambda$6734 ($^2I_{13/2}\rightarrow{^4G}_{11/2}$) &
	$\le4.70$E$-7$ & \nodata & \nodata & \nodata & \nodata & \nodata \\
$\lambda$6740 ($^2I_{11/2}\rightarrow{^4G}_{11/2}$) &
	$\le3.46$E$-7$ & 5.57E$-7$ & 3.36E$-7$ & \nodata & \nodata & \nodata \\
$\lambda$7223 ($^4F_{9/2}\rightarrow{^4D}_{7/2}$) &
	\nodata &\nodata &\nodata &\nodata & \nodata & 5.79E$-8$\tablenotemark{a} \\
\enddata
\tablecomments{The final adopted values and their uncertainties are listed in
Table~\ref{t1}.}
\tablenotetext{a}{These abundances have been derived from lines whose
intensities are highly uncertain \citep{izo01}
and are not used in the calculation of the final values.}
\end{deluxetable}

\clearpage

\begin{deluxetable}{lllllll}
\tabletypesize{\footnotesize}
\tablecolumns{7}
\tablecaption{Physical Conditions and Ionic Abundances: $\{X\}=12+\log(X)$
\label{t1}}
\tablewidth{0pt}
\tablehead{
\colhead{} & \colhead{$T_{\rm e}$\tablenotemark{a}} &
\colhead{$N_{\rm e}$} & \colhead{} & \colhead{} & \colhead{} & \colhead{} \\
\colhead{Object} & \colhead{(K)} & \colhead{(cm$^{-3}$)} &
\colhead{$\{\mbox{O}^{+}/\mbox{H}^+\}$} &
\colhead{$\{\mbox{O}^{++}/\mbox{H}^+\}$} &
\colhead{$\{\mbox{Fe}^{++}/\mbox{H}^+\}$} &
\colhead{$\{\mbox{Fe}^{3+}/\mbox{H}^+\}$} 
}
\startdata
\objectname{30~Doradus}  & $10800^{+350}_{-300}$ & \phn\phn440$\pm190$ &
	$7.56^{+0.05}_{-0.06}$ & \nodata &
	5.22$\pm$0.07 & \nodata \\[0.1cm]
	& 10000$\pm$200 & \phn\phn440$\pm$190 &
	\nodata & 8.25$\pm$0.04 &
	\nodata & $\le5.54$ \\[0.1cm]
\objectname{IC~4846}  & 12200$^{+5000}_{-2000}$ & \phn8700$\pm$3900 &
	$6.99^{+0.31}_{-0.55}$ & \nodata &
	$4.70^{+0.24}_{-0.60}$ & \nodata \\[0.1cm]
	& 10600$^{+900}_{-600}$ & \phn8700$\pm$3900 &
	\nodata & $8.50^{+0.09}_{-0.12}$ &
	\nodata & $5.75^{+0.20}_{-0.34}$ \\[0.1cm]
\objectname{M42}  & 10000$^{+1600}_{-1000}$ & \phn6400$\pm$2800 &
	$7.92^{+0.23}_{-0.32}$ & \nodata &
	$5.52^{+0.16}_{-0.20}$ & \nodata \\[0.1cm]
	& \phn8300$^{+600}_{-400}$ & \phn6400$\pm$2800 &
	\nodata & $8.46^{+0.11}_{-0.15}$ &
	\nodata & $5.52^{+0.24}_{-0.46}$ \\[0.1cm]
\objectname{N88A bar}  & 14200$\pm$400 & 10200$^{+4800}_{-6100}$ &
	$6.96^{+0.11}_{-0.19}$ & 7.97$\pm$0.04 &
	5.23$\pm$0.05 & 5.55$\pm$0.06 \\[0.1cm]
\objectname{N88A sq.~A}  & 13500$^{+900}_{-600}$ & \phn1500$^{+4500}_{-1000}$ &
	$6.69^{+0.19}_{-0.14}$ & $8.02^{+0.06}_{-0.08}$ &
	$5.12^{+0.11}_{-0.13}$ & $5.62^{+0.11}_{-0.16}$ \\[0.1cm]
\objectname{SBS~0335$-$052}  & 20200$^{+800}_{-700}$ &
	\phn\phn300$^{+400}_{-270}$ &
	$5.92^{+0.06}_{-0.07}$ & $7.24^{+0.03}_{-0.04}$ &
	4.51$\pm$0.11 & $5.08^{+0.18}_{-0.32}$ \\
\enddata
\tablenotetext{a}{When two values are given for $T_{\rm e}$, the first one is
that derived from the [\ion{N}{2}] lines; the second value is from
the [\ion{O}{3}] lines. A single entry shows the $T_{\rm e}$ obtained from
the [\ion{O}{3}] lines.}
\end{deluxetable}

\clearpage

\begin{deluxetable}{llllllc}
\tabletypesize{\footnotesize}
\tablecaption{Total Abundances: $\{X\}=12+\log(X)$ \label{t2}}
\tablewidth{0pt}
\tablehead{
\colhead{Object} & \colhead{$\{\mbox{O}/\mbox{H}\}$} &
\colhead{$\{\mbox{Fe}/\mbox{H}\}$\tablenotemark{a}} &
\colhead{$\log\mbox{Fe}/\mbox{O}$\tablenotemark{a}} &
\colhead{$\{\mbox{Fe}/\mbox{H}\}$\tablenotemark{b}} &
\colhead{$\log\mbox{Fe}/\mbox{O}$\tablenotemark{b}} &
\colhead{$\mbox{Fe}^{3+}_{\rm exp}/\mbox{Fe}^{3+}\,$\tablenotemark{c}}
}
\startdata
\objectname{30~Doradus}  & 8.33$\pm$0.03 & 6.04$\pm$0.07 &
	$-2.29^{+0.06}_{-0.07}$ & $\le5.71$ &
	$\le-2.62$ & $\ge2.7$\phn \\[0.1cm]
\objectname{IC~4846}  & 8.51$^{+0.09}_{-0.12}$ & 6.26$^{+0.16}_{-0.25}$ &
	$-2.25^{+0.15}_{-0.20}$ & 5.78$^{+0.19}_{-0.30}$ &
	$-2.73^{+0.20}_{-0.33}$ & 3.2$^{+2.3}_{-2.4}$\\[0.1cm]
\objectname{M42}  & 8.57$^{+0.10}_{-0.13}$ & 6.27$^{+0.19}_{-0.22}$ &
	$-$2.30$\pm$0.15 & 5.86$^{+0.15}_{-0.20}$ &
	$-2.71^{+0.17}_{-0.25}$ & 4.4$^{+4.1}_{-3.9}$\\[0.1cm]
\objectname{N88A bar}  & 8.01$\pm$0.04 & 6.32$^{+0.15}_{-0.10}$ &
	$-1.69^{+0.17}_{-0.10}$ & 5.72$\pm$0.05 &
	$-2.30^{+0.03}_{-0.04}$ & 5.5$^{+2.5}_{-1.3}$\\[0.1cm]
\objectname{N88A sq.~A}  & 8.04$^{+0.06}_{-0.08}$ & 6.51$^{+0.14}_{-0.24}$ &
	$-1.53^{+0.14}_{-0.25}$ & 5.74$^{+0.10}_{-0.13}$ &
	$-2.30^{+0.08}_{-0.09}$ & 7.5$^{+3.2}_{-3.6}$\\[0.1cm]
\objectname{SBS~0335$-$052}  & 7.26$\pm$0.04 & 5.89$^{+0.13}_{-0.14}$ &
	$-$1.37$\pm$0.13 & 5.19$^{+0.15}_{-0.23}$ &
	$-2.07^{+0.15}_{-0.23}$ & 6.1$^{+3.7}_{-3.5}$\\
\enddata
\tablenotetext{a}{$\mbox{Fe}/\mbox{H}=
	1.1\,[(\mbox{Fe}^++\mbox{Fe}^{++})/\mbox{O}^+]\,\mbox{O}/\mbox{H}$.}
\tablenotetext{b}{$\mbox{Fe}/\mbox{H}=
\mbox{Fe}^+/\mbox{H}^++\mbox{Fe}^{++}/\mbox{H}^++\mbox{Fe}^{3+}/\mbox{H}^+$.}
\tablenotetext{c}{The ratio between the expected and calculated values
of $\mbox{Fe}^{3+}/\mbox{H}^+$, where $\mbox{Fe}^{3+}_{\rm exp}/\mbox{H}^+$ is
derived from $\mbox{Fe}^+/\mbox{H}^++
\mbox{Fe}^{++}/\mbox{H}^++\mbox{Fe}^{3+}_{\rm exp}/\mbox{H}^+=
1.1\,[(\mbox{Fe}^++\mbox{Fe}^{++})/\mbox{O}^+]\,\mbox{O}/\mbox{H}$.}
\end{deluxetable}

\end{document}